# Influence of Heat Treatment on the Corrosion Behavior of Purified Magnesium and AZ31 Alloy

Sohrab Khalifeh[a1], and T. David Burleigh[2]

[a] Department of Materials and Metallurgical Engineering Department, New Mexico Institute of Mining and Technology, Socorro, New Mexico, USA

**Abstract**

Magnesium and its alloys are ideal for biodegradable implants due to their biocompatibility and their low stress shielding. However, they can corrode too rapidly in the biological environment. The objective of this research was to develop heat treatments to slow the corrosion of high purified magnesium and AZ31 alloy in simulated body fluid at 37°C. Heat treatments were performed at different temperatures and times. Hydrogen evolution, weight loss, PDP, and EIS methods were used to measure the corrosion rates. Results show that heat treating can increase the corrosion resistance of HP-Mg by 2x and AZ31 by 10x.

## 1. Introduction

Magnesium and its alloys are particularly considered as biomaterials due to their desired properties such as biodegradation, biocompatibility, and low stress shielding [1]. Magnesium can be easily dissolved by the following oxidation and reduction reactions in the biological environment, and produce the hydrogen gas [2].

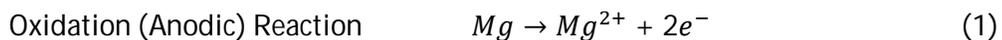
Oxidation (Anodic) Reaction $\quad Mg \rightarrow Mg^{2+} + 2e^-$ $\quad$ (1)

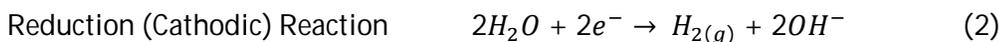
Reduction (Cathodic) Reaction $\quad 2H_2O + 2e^- \rightarrow H_{2(g)} + 2OH^-$ $\quad$ (2)

---

[1] Corresponding author. *Email address:* skhalife@nmt.edu
[2] burleigh@nmt.edu

All implant materials and the elements used to alloy magnesium are not biocompatible. Corrosion of stainless steel can release metallic ions (e.g. $Ni^+$) into the body and consequently cause problems [3]. Many orthopedic implants cause serious problems in patients in the long term such as heavy metal poisoning and chronic pains. In the case of heart vascular stents, reports indicate that the stents can be potentially dangerous for the patients [4, 5]. Late stent thrombosis, in-stent restenosis, and the need for prolonged anti-platelet drugs cause a high risk for using stents for a long time. Removing the implant after healing period can avoid the problems from long-term presence of biomaterials in human body. Retrieving the implants means a second surgery which is detrimental to the part of body that has recently healed. This can lead to chronic pain in many patients. Any surgery will bring further pain, cost, anesthesia, and possibly postoperative complications for patients. However, using bio-absorbable magnesium implants can prevent the problems associated with a second surgery.

Magnesium is potentially an ideal biomaterial because it is non-toxic, non-allergic, non-immunogenic, and non-carcinogenic [6]. Magnesium is an essential element in biological system, and acts as a bridge element in most of the enzyme systems that regulate several biochemical reactions in the human body. These include protein synthesis, osteogenesis, muscle and nerve function, blood glucose control, and blood pressure regulation [7, 8, 9, 10, and 11]. Too much magnesium does not pose a high health risk in human body, because the kidneys eliminate excess magnesium in the urine.

Another distinguishing characteristic of magnesium as a biomaterial is less stress shielding compare to common biomaterials (stainless steel, titanium, hydroxyapatite, etc.) because of minimal difference between the elastic moduli of magnesium (45 GPa) versus bone (40 GPa)(cortical bone/tibia) [12, 13, 14, 15, 16, 17, and 18]. When a stainless steel implant or prosthesis is attached to a bony structure, the implant will carry a larger portion of the load than the skeleton due to the large elastic moduli of the stainless steel (200 GPa) and bone. According to the Wolff's law, the bone mass will decrease, and become weaker [19]. The distinct advantages of magnesium and its alloys make them excellent candidates for biodegradable implants in orthopedic applications.

On the other hand, using magnesium as biodegradable implant has been limited due to the rapid and uncontrollable corrosion that can lead to serious problems such as hydrogen bubbles, caustic burning, loss of mechanical properties, and disappearance of the implant before complete healing [1, 2, 20, 21, 22, and 23].

Many investigations have been conducted to find methods and techniques to decrease the corrosion rate of magnesium. Alloying, coating, and purifying can be used as methods to slow the corrosion rate of biodegradable magnesium implants [24, 25, 26, 27, 28, and 29]. However the alloying, and coating must not degrade the advantages of magnesium such its biodegradation and biocompatibility. These requirements limit the types of alloying and coatings for magnesium. Heat treatment also can have a great influence on corrosion resistance of magnesium and its alloys due to the modification of microstructure [30, 31].

Aung et al. [31] showed that homogenization treatment and ageing treatment can improve the corrosion resistance of AZ91D magnesium alloy in 3.5% NaCl solution.

In this paper, the influence of heat treatment on the microstructure and consequently on the corrosion behavior as a function of temperature and time of heat treatment on purified magnesium and magnesium alloy (AZ31-B) as an *in vitro* test were investigated. Reduction in the corrosion rate of magnesium by heat treatment can be due to the homogenization of the microstructure and subsequently minimizing the intermetallic particles that act as cathodic sites.

## 2. Material and Methods

### 2.1. Composition and Heat Treatments

The chemical composition of the AZ31-B and the high purified magnesium (HP Mg) used in this research are given in Table 1.

Table 1. Composition for HP Mg and commercial AZ31-B used in this work. All compositions are given in weight percent

| Material | Mg    | Al    | Zn    | Ca    | Si    | Mn    | Fe    | Ni    | Cu      | Zr     | Na    |
|----------|-------|-------|-------|-------|-------|-------|-------|-------|---------|--------|-------|
| HP Mg    | 99.97 | 0.002 | 0.005 | 0.001 | 0.014 | 0.001 | 0.003 | 0.002 | >0.0002 | >0.002 | 0.002 |
| AZ31-D   | Bal.  | 2.953 | 0.687 | >0.01 | 0.031 | 0.469 | 0.001 | 0.001 | 0.013   | -      | -     |

The AZ31 (12.7 mm in diameter) and HP Mg (18 mm in diameter) rods were cut into 10 mm long cylinders. The heat treatments (T1-T12) were performed at different temperature (350, 400, and 450°C) for 3, 6, 12, and 24 hours under protective argon gas followed by furnace cooling to 400°C, then water quenched to room temperature. Table 2 shows the time and temperatures for the T1-T12 heat treatments.

Table 2. Solution treatments.

|        | 3 h | 6 h | 12 h | 24 h |
|--------|-----|-----|------|------|
| 350°C  | T1  | T2  | T3   | T4   |
| 400°C  | T5  | T6  | T7   | T8   |
| 450°C  | T9  | T10 | T11  | T12  |

Different techniques were used to investigate the corrosion behavior of heat treated AZ31, and HP Mg. Weight loss, hydrogen evolution, electrochemical impedance spectroscopy (EIS), and potentiodynamic polarization (PDP) were all used in this research. The heat treated samples were cold mounted in epoxy with a 1.27 cm$^2$ (AZ31), and 2.54 cm$^2$ (HP Mg) surface area. Next, the specimen surfaces were grounded with sand paper (600-1200 grit), and then polished to 1 $\mu$m oil-based diamond slurry, degreased with ethanol, washed with deionized water, and dried using compressed air.

### 2.2. Electrolyte

All corrosion tests were performed as *in vitro* based on the human biological environment. The samples were immersed in simulated body fluid (SBF) at 37°C. The SBF was made according to protocol for preparing SBF at 36.5°C and pH 7.4 [32]. Ion concentrations of SBF versus human blood plasma are shown in Table 3.

*Table 3. SBF versus human Blood plasma [32]*

|  | Simulated Body Fluid (SBF)(mM) | Blood Plasma (mM) |
|---|---|---|
| Na+ | 142.0 | 142.0 |
| K+ | 5.0 | 5.0 |
| Mg+ | 1.5 | 1.5 |
| Ca+ | 2.5 | 2.5 |
| Cl- | 148.8 | 103.2 |
| HCO3- | 4.2 | 27.0 |
| HPO4 2- | 1.0 | 1.0 |
| SO4 -2 | 0.5 | 0.5 |

### 2.3. Hydrogen Evolution Tests

Three separate specimens were used for each heat treatment condition and test. Hydrogen evolution was based on the released hydrogen gas (mL) due to magnesium oxidation (equation 3). The released hydrogen volume rate is defined by the following equation:

$$V_{H_2} = \frac{\Delta v_{H_2}}{At} \qquad (3)$$

Where $t$ is time of immersion (day), and $A$ is the surface area that is exposed to the immersion solution $A$ (cm²). The corrosion rate in mm. year$^{-1}$ is determined as $C.R._{H_2}$ by the following calculation:

$$C.R._{H_2} = \left(\frac{\Delta v_{H_2}}{At}\right)\left(\frac{mL_{H_2}}{cm^2 \cdot day}\right)\left(\frac{10^{-3}L}{mL}\right)\left(\frac{0.089 g_{H_2}}{L_{H_2}}\right)\left(\frac{mol\ H_2}{2 \times 1.008 g_{H_2}}\right)$$

$$\left(\frac{mol\ Mg}{mol\ H_2}\right)\left(\frac{24.305 g Mg}{mol\ Mg}\right)\left(\frac{cm^3}{1.738 Mg}\right)\left(\frac{10mm}{cm}\right)\left(\frac{365\ day}{year}\right) \qquad (4)$$

$$C.R._{H_2} = 2.27\ V_{H_2}\ (mm.year^{-1}) \qquad (5)$$

The advantages of the hydrogen evolution test are the estimation of instantaneous corrosion rate during the test, smaller theoretical and experimental errors, and set up is also quick and easy.

### 2.4. Weight Loss Test

Weight-loss is the second method and is the classic and easiest method to measure the corrosion rate. In the weight-loss method, each specimen is degreased by ethanol, washed with deionized water, dried with compressed air, weighted before the immersion test, and documented as $W_{before}$. After the immersion test, the corrosion products were removed by dropping concentrated nitric acid on the surface of the specimen for 10 seconds. Then, the specimens are washed with deionized water, dried by compressed air, and weighted to obtain $W_{after}$. The weight loss rate, $W_L$ ($mg \cdot cm^{-2} \cdot day^{-1}$) was determined according to the following equation;

$$W_L = \frac{W_{before} - W_{after}}{At} \qquad (6)$$

Where $t$ is time of immersion (days), and A is the surface area that was exposed to the immersion solution A (cm²). The corrosion rate in mm·$year^{-1}$ was determined as $C.R._{WL}$ by the following calculation:

$$C.R._{WL} = \left(\frac{W_{before} - W_{after}}{At}\right)\left(\frac{mg}{cm^2 \cdot day}\right)\left(\frac{cm^3}{1.738g}\right)\left(\frac{10^{-3}g}{mg}\right)\left(\frac{10mm}{cm}\right)\left(\frac{365\ day}{year}\right) \qquad (7)$$

$$C.R._{WL} = \left(\frac{W_{before} - W_{after}}{At}\right)(2.1)\left(\frac{mm}{year}\right) = 2.1\ W_L(mm \cdot year^{-1}) \qquad (8)$$

The weight-loss method can have errors in $W_{after}$, and subsequently in $C.R._{WL}$, because of difficulties removing the corrosion product of magnesium. The results from measuring the $W_{before}$ and $W_{after}$ by dropping concentrated nitric acid on the surface (1200 grit) of the as-received non-corroded HP Mg and AZ31 specimens for 10 second showed that the HP Mg dissolution (1.1 mg) was higher than the AZ31 (0.1 mg). The HP Mg dissolution corresponded to increasing the corrosion rate by 1%.

The absorption of water by the epoxy was another possible error. In this case, the bare epoxy cylinder was immersed in SBF in a same period of weight-loss method. Then, the bare epoxy was weighed continuously for 20 weeks. The results show that the epoxy could absorb a few tenths of a weight percent water, and loose the water also by evaporation. However this effect would be the same for all samples, so the relative results were the same. In future studies, it would be recommended not to use epoxy mounted samples for the weight loss test.

## 2.5. Electrochemical Measurements

Electrochemical impedance spectroscopy (EIS), and potentiodynamic polarization (PDP) of the specimens were measured in SBF at 37°C using a PARSTAT 2263 potentiostat. For all measurements, a three-electrode electrochemical cell was used, with a KCl saturated calomel electrode (SCE) as a reference electrode and platinum wire as a counter electrode. The specimens were polished up to 1200 grit, followed by washing in deionized water and then dried by compressed air. At least three tests were performed for each condition with three different specimens to confirm the reproducibility of the EIS and PDP measurements. EIS test began after the specimens were immersed in SBF at 37°C, and run in the frequency range from 100 kHz to 100 mHz.

The PDP test was run immediately after the final EIS test. The initial potential was -250 mV relative to open circuit potential and stopped at +1600 mV versus SCE scan at a rate of 10 mV/s. Extrapolation of the cathodic and anodic region of Tafel behavior gave the corrosion rate, $i_{corr\ (A/cm^2)}$ at $E_{corr}$.

Polarization resistance, $R_p$ was determined from the EIS Bode plot. The impedance at 100 mHz ($R_p + R_s$) was reduced by the solution resistance at 100 kHz ($R_s$) to obtain $R_p$ (polarization resistance). Then, $R_p$ was converted to the corrosion rate, $i_{corr\ (A/cm^2)}$ by the following equation;

$$R_p = \frac{\beta_a \beta_c}{2.3\ i_{corr}(\beta_a + \beta_c)} = \frac{B}{i_{corr}} \qquad (9)$$

Where $\beta_a$ and $\beta_c$ are anodic and cathodic Tafel constants, respectively, and $B$ is proportionality constant, which was determined from PDP electrochemical data. $i_{corr\ (A/cm^2)}$. Each EIS and PDP test was converted to the corrosion rate (mm/year) as a $C.R._{EIS/PDP}$ by the following calculation;

$$C.R._{EIS/PDP} = i_{corr}\left(\frac{A}{cm^2}\right)\left(\frac{C}{A\cdot sec}\right)\left(\frac{e^{-1}}{1.61\times 10^{-19}C}\right)\left(\frac{Mg\ atom}{2e^-}\right)$$

$$\left(\frac{mol\ Mg}{6.023\times 10^{23}\ Mg\ atom}\right)\left(\frac{24.305g\ Mg}{mol\ Mg}\right)\left(\frac{cm^3}{1.738g}\right) \qquad (10)$$

$$\left(\frac{10mm}{cm}\right)\left(\frac{3600sec}{h}\right)\left(\frac{24h}{day}\right)\left(\frac{365day}{year}\right)$$

$$= 22.74\times 10^3\ i_{corr}\ mm\cdot year^{-1}$$

### 2.6. Microstructure and Morphology Characterization

The microstructure of each specimen of each heat treatment cycle was investigated by optical microscopy (Olympus GX41 inverted microscope), and Hirox (KH-7700 digital microscope). The heat treated samples were polished up to 0.05 $\mu$m alumina slurry solution, then etched with acetic glycol (20 mL acetic acid, 1 mL HNO$_3$, 60 mL ethylene glycol, and 20 mL water) for 10 seconds [33]. An electron microprobe (Cameca SX-100) was used in order to analysis the distribution of secondary phases, alloying elements, and impurities as a cathodic sites before and after the heat treatment.

## 3. Results

### 3.1. Microstructure

Figure 1 illustrates the microstructure of the different heat treated AZ31 samples. Figure 1a shows the microstructure of as-received AZ31. Figures 1(b-m) show the heat treated T1-T12 respectively. As can be seen in Figure 1, there is a small increase in the average grain diameter of AZ31 for longer time and higher temperature. For example, the average grain diameter of as-received AZ31 increased from 26 $\mu$m to 43 $\mu$m by T12 heat treatment. For AZ31, it seems that alloying elements (Al, Zn, or Mn), and $\beta$ precipitates can pin the grain boundaries, and prevent movement at the low temperature heat treatments (≤450°C).

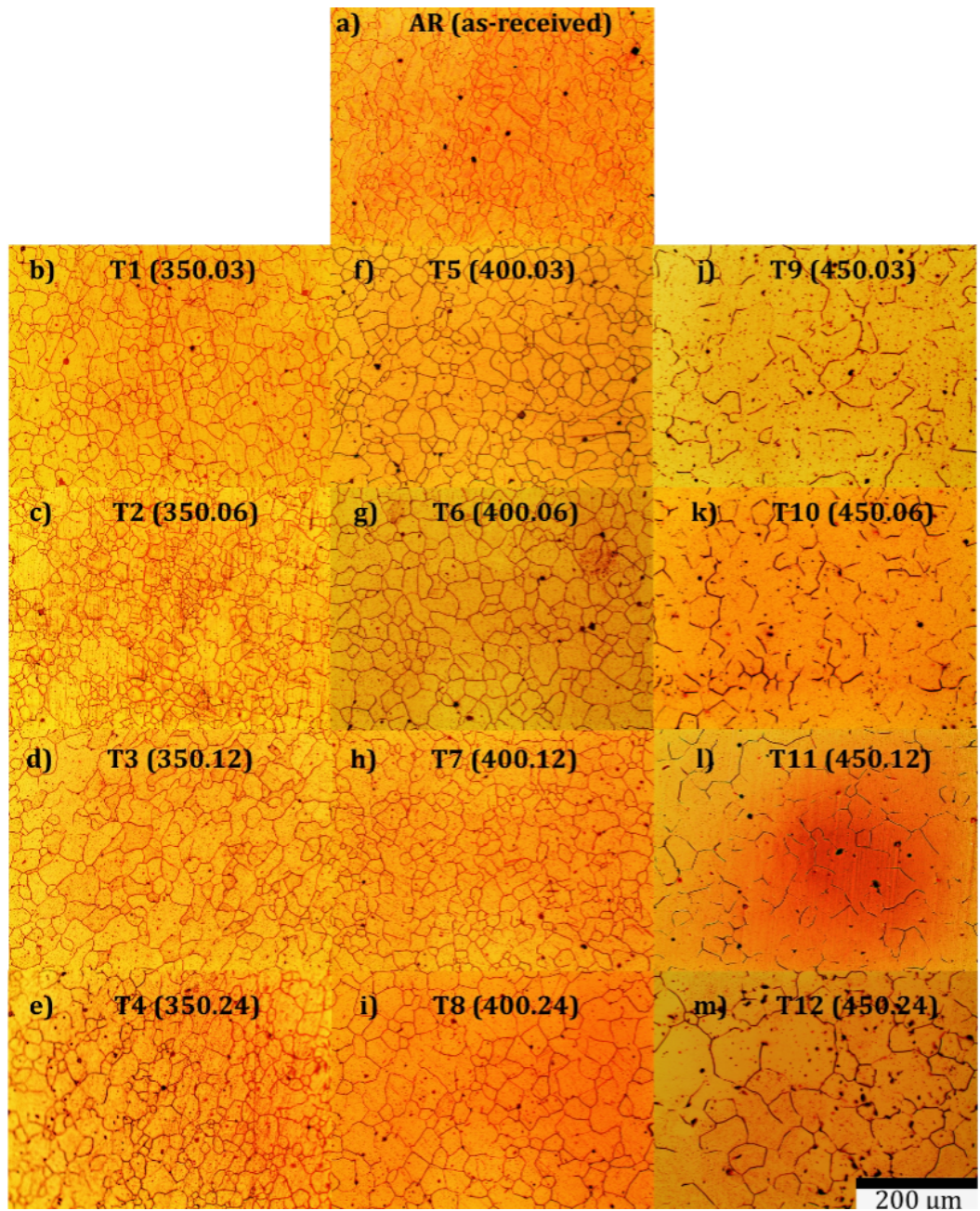

Figure 1. The microstructure of different heat treated AZ31 samples by optical microscopy.

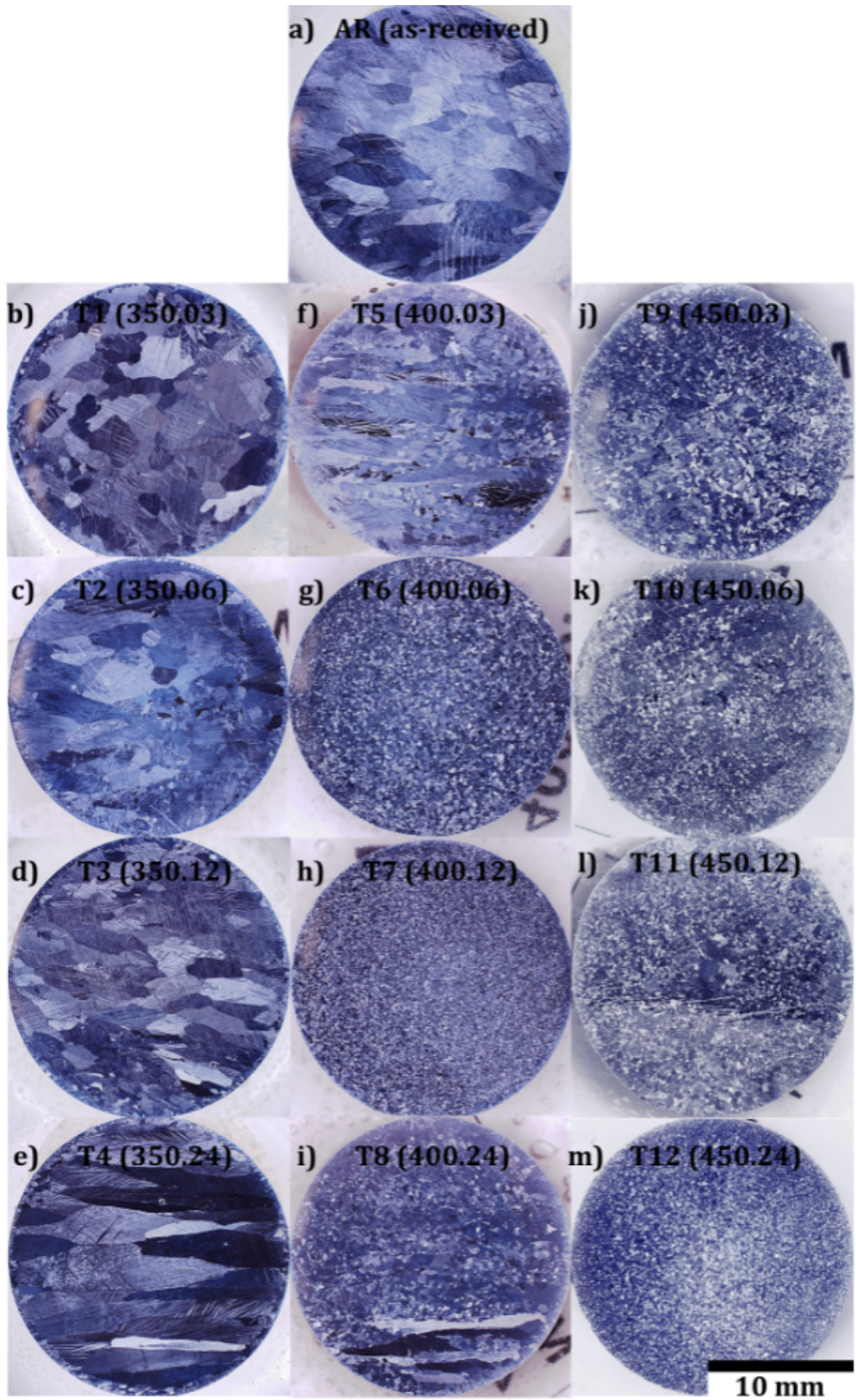

Figure 2. The images for as-received and heat treated HP Mg.

However, the grain size can be severely changed by heat treatment of HP Mg. In Figure 2, the grains can grow up to the length of 15 mm (T4). Figures 3 and 4 show the elemental maps (of analysis) (Al, Zn, and Mn), and the backscatter electron images of the as-received and heat treated AZ31, and HP Mg by the electron microprobe. The effect of heat treatment on the size of precipitates and secondary phases clearly are seen in the Figures 3 (a, and e). The zones shaded red represent the precipitate free zones. The microstructure of heat treated magnesium alloy has been made more uniform compared to the as-received magnesium alloy.

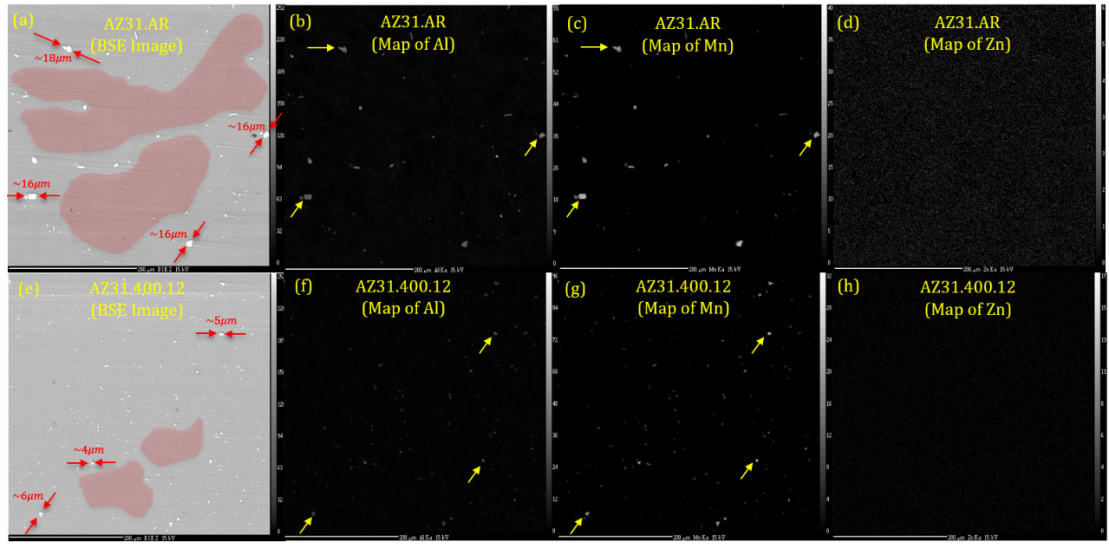

*Figure 3. The images from electron micro-probe for AZ31, (a) BSE image of as-received AZ31. Red arrows show the precipitates, and red regions show the area without precipitates, (b) map of Al (as-received AZ31), (c) map of Mn (as-received AZ31), (d) map of Zn (as-received AZ31), and (e) BSE image of heat treated AZ31, T7. Red arrows show the precipitates, and red regions show the area without precipitates, (f) map of Al (heat treated AZ31, T7), (g) map of Mn (heat treated AZ31, T7), and (h) map of Zn (heat treated AZ31, T7).*

The map of zinc (Figure 3 d, and h) shows that zinc was not detectable. The maps of Al and Mn (Figure 3 b, c, f, and g) illustrate that most of the precipitates in AZ31 contain Al and Mn. These precipitates (yellow arrows) are most likely MnAl, $MnAl_4$, and $MnAl_6$ in the as-received AZ31, but transform to $MnAl_6$ during the heat treatment [17]. Figure 4 shows the map of Ni, Si, and BSE images for as-received and heat treated HP Mg. The map of Ni and Si (Figure 4 b, c, e, and f) show that the impurities have been dissolved by the heat treatment, which can improve the corrosion resistance of HP Mg.

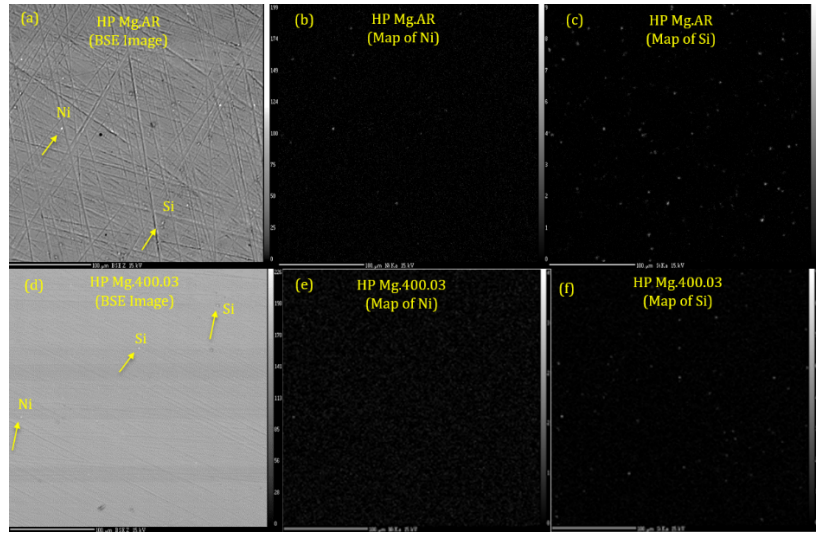

*Figure 4. The images from electron micro-probe for HP Mg, (a) BSE image of as-received HP Mg. Yellow arrows show the Ni and Si precipitates, (b) map of Ni (as-received HP Mg), (c) map of Si (as-received HP Mg), (d) BSE image of heat treated HP Mg, T5, (e) map of Ni (heat treated HP Mg, T5), and (f) map of Si (heat treated HP Mg, T5).*

### 3.2. Hydrogen Evolution and Weight Loss Test Results

The corrosion rate of the immersed as-received and heat treated AZ31 in the simulated body fluid at 37°C based on hydrogen evolution test and weight loss test are illustrated in Figure 5 and 6, respectively. The minimum corrosion rate occurs for T7 (at 400°C for 12 h). T1 has the least influence on corrosion resistance of AZ31. It means, heating up to 350°C just for 3 hours, has the smallest improvement on the corrosion rate of AZ31. The corrosion rate was reduced by increasing the holding time for the heat treatment processes at 350°C (Figures 5 and 6). Also, the corrosion rates for heat treated AZ31 at 24 hours for all three temperature groups (350, 400, and 450°C) are very close together. This shows that the predominant factor at 350°C is holding time, and the predominant factor for short time (3 h) of heat treating is temperature.

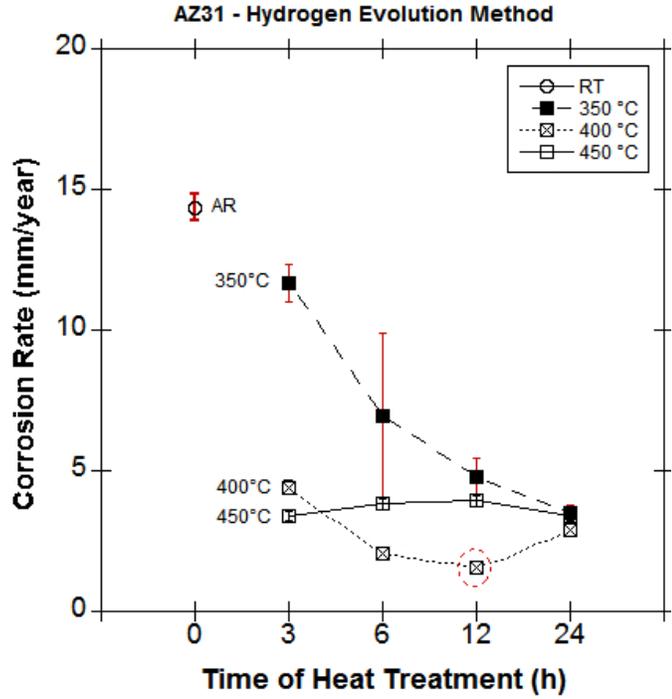

*Figure 5. Effect of time and temperature of heat treatment on corrosion rate of AZ31 based on hydrogen evolution test.*

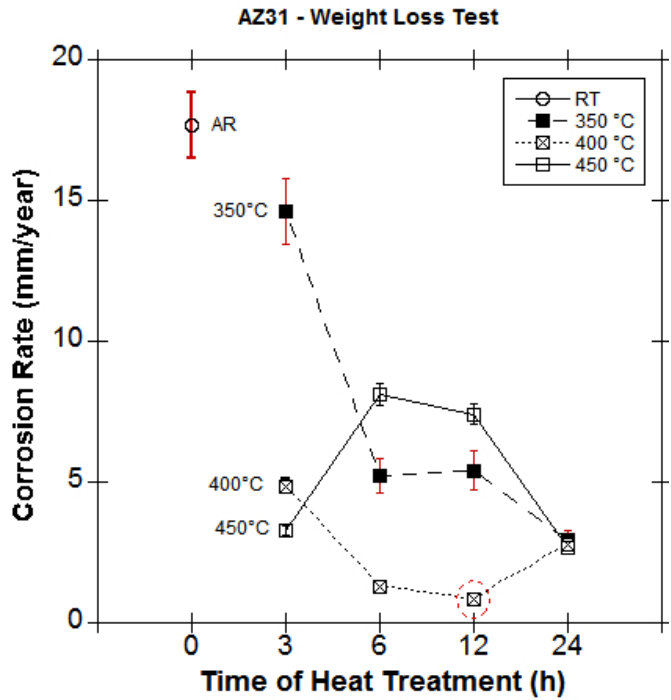

*Figure 6. Effect of time and temperature of heat treatment on corrosion rate of AZ31 based on weight loss test.*

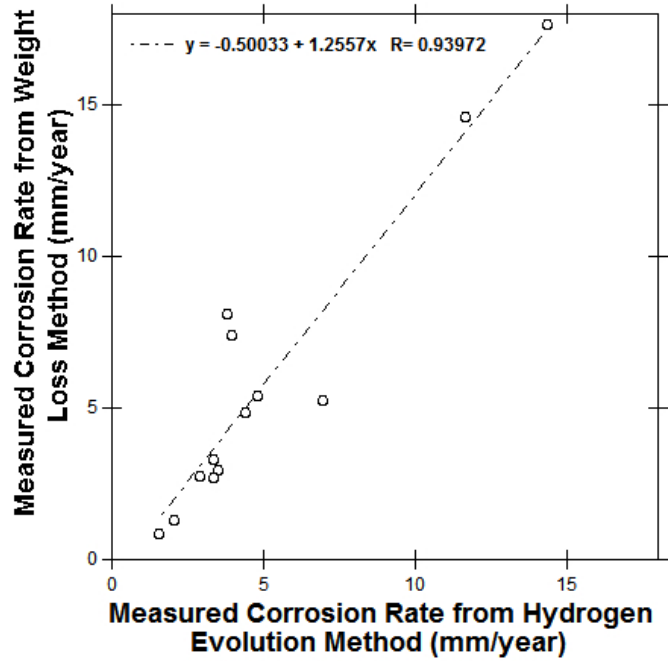

*Figure 7. Regression line from the corrosion rate (mm/year) of AZ31 based on both hydrogen evolution and PDP methods.*

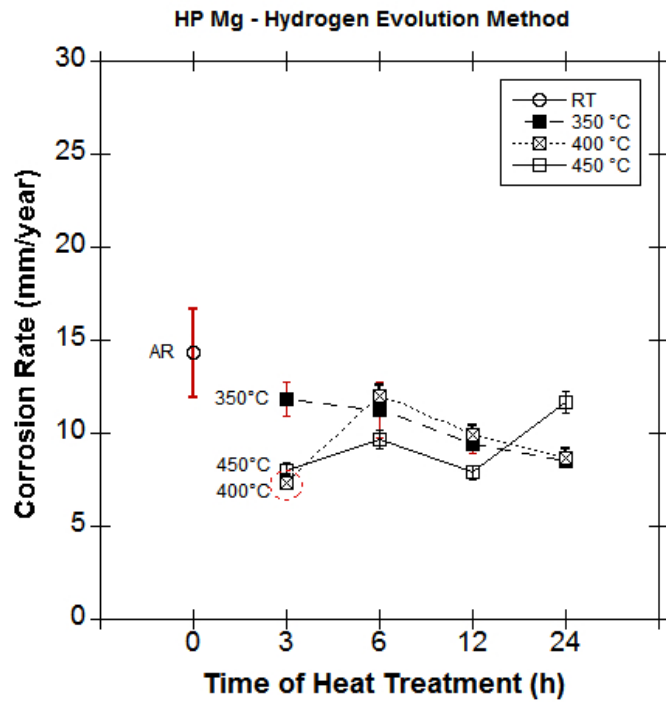

*Figure 8. Effect of time and Temperature of heat treatment on corrosion rate of HP Mg based on hydrogen evolution method.*

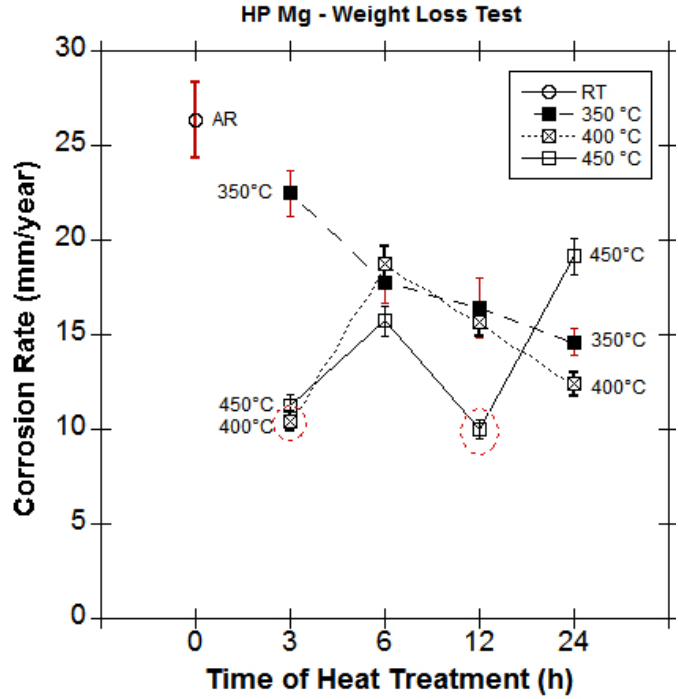

*Figure 9. Effect of time and Temperature of heat treatment on corrosion rate of HP Mg based on weight loss method.*

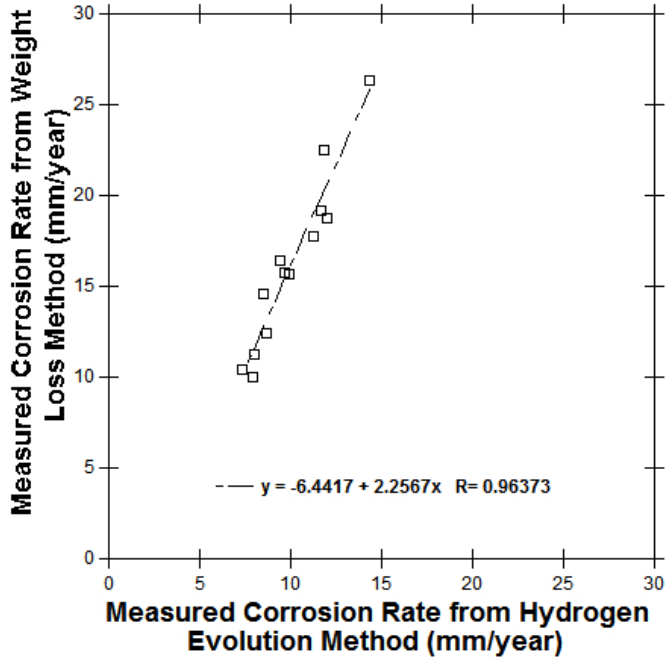

*Figure 10. Regression line from the corrosion rate (mm/year) of HP Mg based on both hydrogen evolution and weight loss methods.*

Weight loss test results confirm the results come from the hydrogen evolution method (Figure 6). Figure 7 shows the regression line from the hydrogen evolution and weight loss results for AZ31. The regression line has a slope of 1.25, and the regression coefficient R=0.94 which means that the results from both methods are reliable. In addition, both methods illustrate that the minimum corrosion rate for AZ31 is at T7.

In the same way, the corrosion rate of the immersed as-received and heat treated HP Mg (3N) based on hydrogen evolution test and weight loss methods are illustrated in Figures 8 and 9, respectively. The optimum results for HP Mg (Figure 8) was obtained at T5 (400°C, for 3h) based on the hydrogen evolution method. As can be seen in the Figure 8, the corrosion rate for 350°C group gradually dropped with increasing the time of heat treatment. But the 450°C group shows different behavior, and with increasing the time of heat treatment, the corrosion rate increased.

Figure 9 shows the weight loss test results of HP Mg, which agree with the results from hydrogen evolution method. T5 heat treatment leads to one of the lowest corrosion rate along with the T11 (450°C, for 12 h), which are marked by dashed red circles. The corrosion rates from weight loss method are generally higher than the corrosion rates from hydrogen evolution method. The regression line for HP Mg (Figure 10) has a slope value of 2.26.

### 3.3. PDP and EIS Results

In order to validate the corrosion rates of AZ31 and HP Mg provided by hydrogen evolution and weight loss methods, PDP and EIS test were performed to as-received and heat treated AZ31 and HP Mg. Figure 11 and 12 illustrate the corrosion rate (mm/year) of as-received and heat treated AZ31 based on the PDP and EIS tests, respectively. The results from both tests confirm that the minimum corrosion rate is obtained by T7 heat treatment. In addition, with increasing the time and temperature of heat treatment, the corrosion resistance of AZ31 was improved.

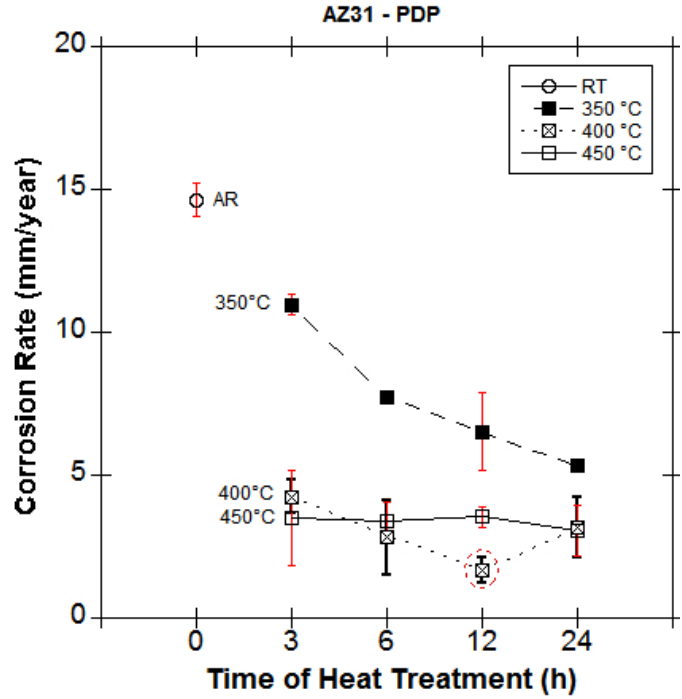

*Figure 11. Effect of time and temperature of heat treatment on corrosion rate of AZ31 based on PDP.*

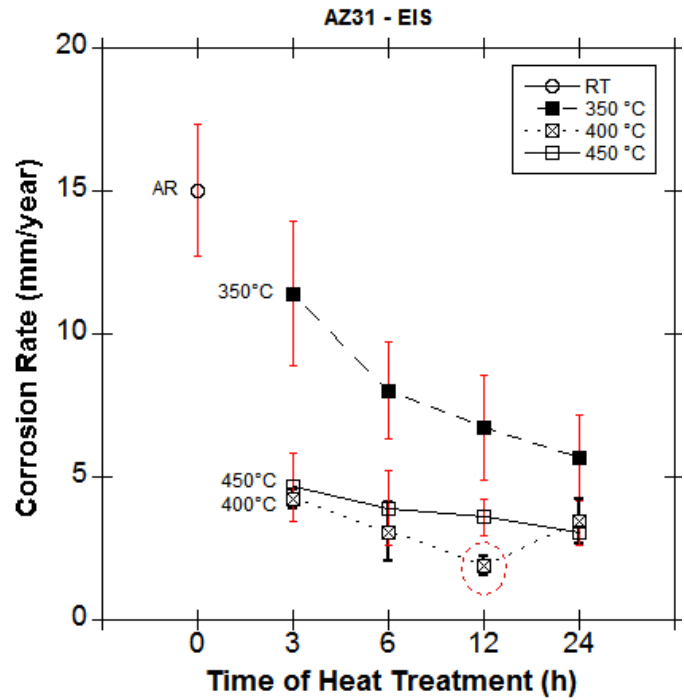

*Figure 12. Effect of time and temperature of heat treatment on corrosion rate of AZ31 based on EIS.*

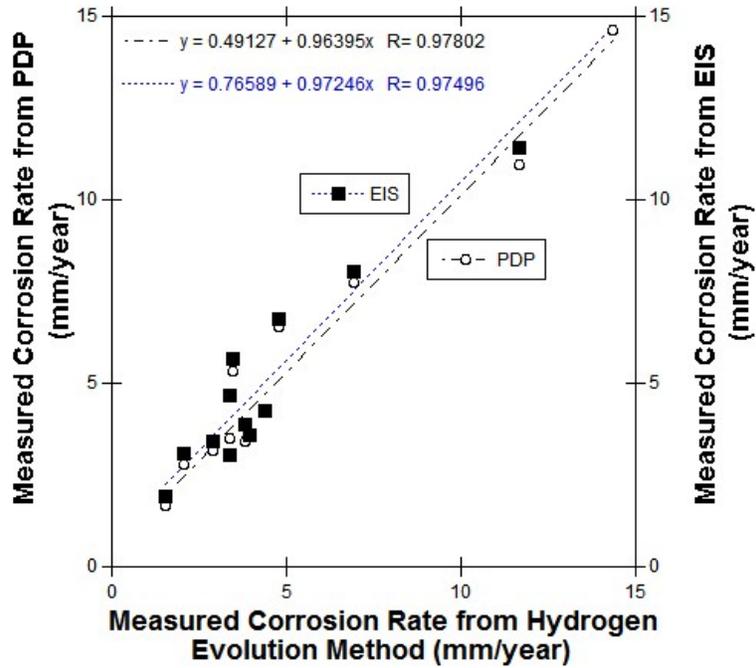

*Figure 13. Regression line from the corrosion rate (mm/year) of AZ31 based on hydrogen evolution, PDP (dashed black line), and EIS (dashed blue line) methods.*

The regression line from PDP, EIS, and hydrogen evolution results are shown in Figure 13, and illustrates that all four methods are reliable to determine the corrosion rate of AZ31. The slope values are 0.96 and 0.97 (hydrogen evolution results versus PDP and EIS, respectively) are close to the ideal value 1, and regression coefficient R=0.97, show that the corrosion rates of AZ31 measured by all methods are accurate in this research.

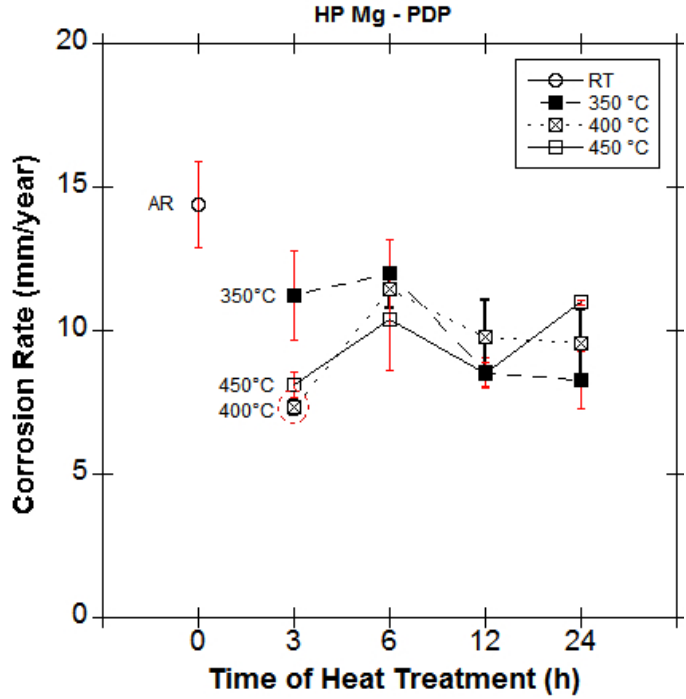

*Figure 14. Effect of time and temperature of heat treatment on corrosion rate of HP Mg based on PDP.*

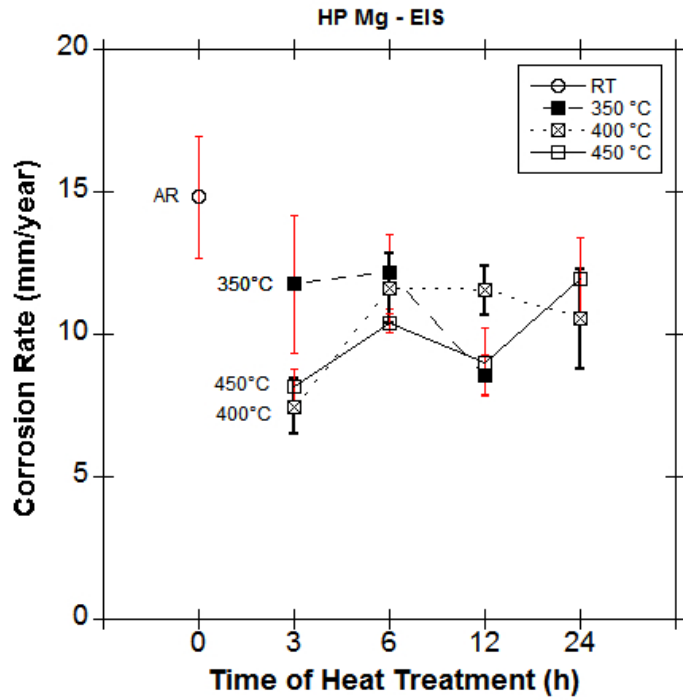

*Figure 15. Effect of time and temperature of heat treatment on corrosion rate of HP Mg based on EIS.*

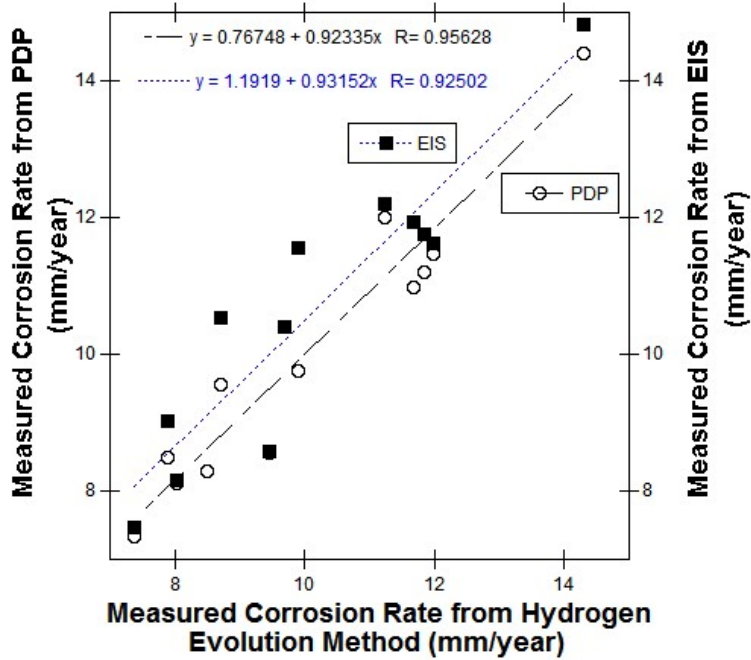

*Figure 16. Regression line from the corrosion rate (mm/year) of AZ31 based on hydrogen evolution, PDP (dashed black line), and EIS (dashed blue line) methods.*

Figure 14 and 15 show the corrosion rate of as-received and heat treated HP Mg based on the PDP and EIS tests versus the holding time of heat treatment at 350, 400, and 450°C, respectively.

The results from electrochemical measurements show that the weight loss method to measure the corrosion rate of HP Mg simply was too high. This could be due to the error caused by the dissolution of HP Mg by nitric acid or the absorption and evaporation of water from the epoxy. The regression line of PDP and EIS results confirm the obtained results by hydrogen evolution rate (Figure 16) by the slope (0.93 and 0.92) close to the ideal value of 1.

## 4. Discussion

In orthopedic implants, biodegradable magnesium implants are considered as an appropriate alternative compared to permanent metallic implants due to no second surgery for removal and less stress shielding. On the other hand, the high corrosion rate of magnesium restricts the use of biodegradable magnesium implant. We have focused on slowing the corrosion rate of magnesium by heat treatment, which does not alter the composition of magnesium. We demonstrated that heat treatment can reduce the corrosion rate by 90% and 50% for as-received AZ31 and HP Mg, respectively. Four different methods were performed to measure the corrosion rate of the AZ31 and HP Mg. All methods confirm that the optimum heat treatment for AZ31 is T7 (at 400°C for 12 h) (Figure 13). All the corrosion rates of all different heat treatments (T1-T12) were similar value, which show that the results and the methods to measure the corrosion rate that are used in this research were reliable.

The results (Figure 5, 6, 11, and 12) illustrate that all heat treatments were effective on reducing the corrosion of biodegradable magnesium. The least impact was obtained by T1 (at 350°C for 3 h), and the corrosion rate was reduced just by 20 percent. The corrosion rates of heat treated AZ31 at 350°C were improved by time of heat treatment. The corrosion rate dropped by 70% compared to as-received AZ31 with 24 hour at 350°C.

Even, a trace amount of heavy metals (Fe, Ni, Co, and Cu) can form as second phase due to the low solubility of these metals in magnesium. The second phase can act as a cathodic site, and can lead to make micro-galvanic corrosion. Therefore, in the presence of heavy metals, magnesium quickly dissolves as a sacrificial anode. Heat treatments can homogenize the microstructure, dissolve the second phases, and subsequently can cause less intermetallic sites. Minimizing the intermetallic sites can lead to less micro-galvanic corrosion between the magnesium matrix as an anode and intermetallic sites as a cathodic site. Changing the size of the second phases is controlled by diffusion mechanism. Temperature and time are two important factors in diffusion mechanism. The results obtained by T1-T4 heat treatments show that 350°C was not high enough to increase the diffusion rate, but with increasing the holding time up to 24 hours, atoms have enough time to diffuse and reduce the size of intermetallic sites.

The microscopic investigations (Figure 1 b-c) do not show any significant change in grain size of AZ31 at 350°C heat treatments. Increasing the heat treatment temperature to 400 and 450°C show lower corrosion rates of AZ31. The microscopic images (Figure 1 j-m) illustrate significant grain growth. Therefore, it seems that the optimum heat treatment temperature is 400°C.

The results from all corrosion methods confirm that the resistance of the heat treated AZ31 at T5-T8 (400°C) have been improved better than heat treated specimens at 350, and 450°C. And also, the optimum time of heat treatment was reached at 12 hours. The microscopic images of the electron microprobe show that the size of the second phases ($MnAl_6$) is dropped 70% on average by T7 heat treatment.

## 5. Conclusions

The influence of heat treatment on the corrosion rate of HP Mg and AZ31 was investigated by different *in vitro* tests. In general, heat treatment showed as an effective role of improving the corrosion rate of HP Mg and AZ31. The results affirm the following:

The corrosion rate of AZ31 was dropped from 15 mm/year down to 1.5 mm/year by heat treating at 400°C for 12 hours (T7).

The size of the intermetallic sites were decreased 70% by T7 heat treatment.

The corrosion rate of HP Mg was reduced from 17 mm/year down to 8.3 mm/year by heat treating at 400°C for 3 hours (T5).

Some impurities such as Ni and Si were dissolved and not detected by T5 heat treating the HP Mg.

The hydrogen evolution test was a more reliable method to measure the corrosion rate of HP Mg and AZ31 alloy.